\documentclass{vietnam}

\def\Journal#1#2#3#4{{#1} {\bf #2}, #3 (#4)}

\def\PLB{{\em Phys. Lett.}  B}

\def\PRD{{\em Phys. Rev.} D}

\def\mco{\multicolumn}

\def\ra{\rightarrow}

\def\ko{K^0}

\def\be{\begin{equation}}
\def\ee{\end{equation}}
\def\bea{\begin{eqnarray}}
\def\eea{\end{eqnarray}}

\usepackage{comment}

\newcommand{\Photo}{}

\begin{document}
\vspace*{4cm}
\title{Charged Lepton Flavor Violating Experiments with Muons}
\author{D. Palo, on behalf of the Mu2e Collaboration}
\address{Fermi National Accelerator Laboratory,
Batavia Illinois,
60542,
USA}

\maketitle\abstracts{
We report on the status of charged lepton flavor violating (CLFV) experiments with muons. We focus on the three "golden channels": $\mu^{+} \rightarrow e^{+} \gamma$, $\mu^{+} \rightarrow e^{+} e^{-} e^{+}$ and $\mu^{-} N \rightarrow e^{-} N$. The collection of upcoming experiments aim for sensitivity improvements up to $10^{4}$ with respect to previous searches. The MEG II experiment, searching for $\mu^{+} \rightarrow e^{+} \gamma$, is currently in its 4th year of physics data-taking with a published result from its first year of data. The Mu3e experiment is an upcoming experiment searching for $\mu^{+} \rightarrow e^{+} e^{-} e^{+}$ with plans of physics data-taking as soon as 2025. The Mu2e and COMET experiments are upcoming searches for $\mu^{-} N \rightarrow e^{-} N$ with the goal of physics data-taking starting in 2027 and 2026 respectively. This proceeding summarizes the signal signature, expected background, resolutions, and timelines for the mentioned searches.}

\section{Introduction}

\vspace{-0.3cm}
\subsection{Theoretical Introduction\vspace{-0.3cm}}
In the Standard Model with massless neutrinos, there exists inherent symmetries that imply the conservation of the three lepton flavors ($L_{e}, L_{\mu}, L_{\tau}$) and total lepton flavor ($L_{l}=L_{e}+ L_{\mu}+L_{\tau}$) in all Standard Model interactions. Neutrinos are now well understood to have mass, which implies neutrino oscillation. This oscillation has been observed by many experiments, which shows that the conservation of lepton flavor is only approximate. 

However, in contrast to neutrinos there has been no observation of lepton flavor violation with charged leptons ($e, \mu, \tau$). Neutrino mixing results in charged lepton flavor violation (CLFV) in the Standard Model (e.g. $\mu^{+} \rightarrow e^{+} \gamma$), but it occurs at a negligible branching fraction due to the small neutrino mass with respect to the mass of the W boson: $\mathcal{B} \propto [\frac{(\Delta m_{\nu})^{2}}{m_{W}^{2}}]^{2}  \sim 10^{-54}$. This branching fraction is so small that it could never be detected by a conceivable experiment. Therefore its observation would imply physics beyond the Standard Model (BSM), making it excellent probe into BSM physics. In particular, there are three "golden" channels of CLFV with muons, which are sensitive to a wide variety of physics BSM: $\mu^{+} \rightarrow e^{+} \gamma$, $\mu^{+} \rightarrow e^{+} e^{-} e^{+}$ and $\mu^{-} N \rightarrow e^{-} N$. This proceedings will focus on these three searches.

\vspace{-0.3cm}
\subsection{Experimental Introduction}
All past CLFV experiments have resulted in a null signal, but current physics models (e.g. SUSY) predict CLFV at a rate that is potentially observable by the modern day experiments. The new suite of experiments plan to improve upon the predecessor experiments by up to a factor of $10^{4}$ in sensitivity. This improvement is achieved via new experimental designs, operation with unprecedented muon beam rates of the order $10^{9} \mu/s$, and detectors designed for precise momentum measurements (e.g. $\sim 150$ keV/$106$ MeV, $\sim 90$ keV/$53$ MeV). 

 There is no true Standard Model background (negligible branching fraction), but there are Standard Model processes that will mimic the signal. The experiments are largely designed around suppressing this background. There are a few background types common to the set of experiments. First, Standard Model muon decays can result in observable particles (electrons, positrons, photons) close to the signal kinematic signature (e.g. momentum) when the muon decay additionally results in non-detected neutrinos carrying off negligible momentum. This is reduced by precise momentum or energy measurements, a common feature in all mentioned experiments. The second type of background is a time coincidence between 2 or more Standard Model processes. These time coincidences are reduced by precise timing resolution and precise vertex resolution at the target. Precise kinematic resolutions are required to achieve the signal to background ratio required for the desired sensitivity; this resolution requirement will be repeated throughout this proceedings. 

In addition to the background suppression, the experiments are designed to take full advantage of the high muon beam rates currently available at PSI, Fermilab and J-PARC ($10^{9} \mu/s$). The detectors, trigger, and electronics all must be capable of handling the high beam rate. Further, the time-coincident background will scale with the beam rate; therefore one must tune the beam rate with the expected kinematic resolution to avoid high background rates. 
To summarize, the muon-based CLFV experiments must make precise kinematic measurements of the muon decay products, and must be capable of handling the high muon beam rates. 

\section{$\mu^{+} \rightarrow e^{+} \gamma$}
\vspace{-0.3cm}
\subsection{Experimental History and Timeline}
The MEG experiment's search for $\mu^{+} \rightarrow e^{+} \gamma$ with its full dataset achieved the lowest sensitivity of any charged lepton flavor violating process (limit of $4.2 \cdot 10^{-13}$ at the 90\% confidence level \cite{MEGFull}). The upgraded experiment, the MEG II experiment, aims to improve upon the limit set by MEG by an order of magnitude \cite{MEGIIDesign}. The experiment published a physics result based on its first year of physics data \cite{MEGIIPhysics} and is now in its 4th year of data-taking with plans of continuing collecting data until PSI's upcoming shutdown.

\vspace{-0.3cm}
\subsection{Signal and Background}
The signal signature is a $e^{+} \gamma$ pair consistent with a $\mu^{+}$ decay at rest in the stopping target. That is, the decay products are time coincident, originate at the same target position and are back-to-back each carrying half the muon rest mass ($52.83 $ MeV). 

The dominant background is the result of a time-coincidence between a positron photon pair from two independent Standard Model processes; the positron is the result of a Michel decay ($\mu^{+} \rightarrow e^{+} \nu_{e} \bar{\nu_{\mu}}$) and the photon is from a radiative muon decay ($\mu^{+} \rightarrow e^{+} \nu_{e} \bar{\nu_{\mu}} \gamma$) or an annihilation in flight ($e^{+} e^{-} \rightarrow \gamma \gamma$). Additional background comes from true radiative muon decays where the neutrinos carry off negligible momentum, thus the detected positron photon pair nearly mimics the signal. 

\vspace{-0.3cm}
\subsection{Experimental Design}
\begin{figure}
\begin{minipage}{0.45\linewidth}
\centerline{\includegraphics[width=7cm]{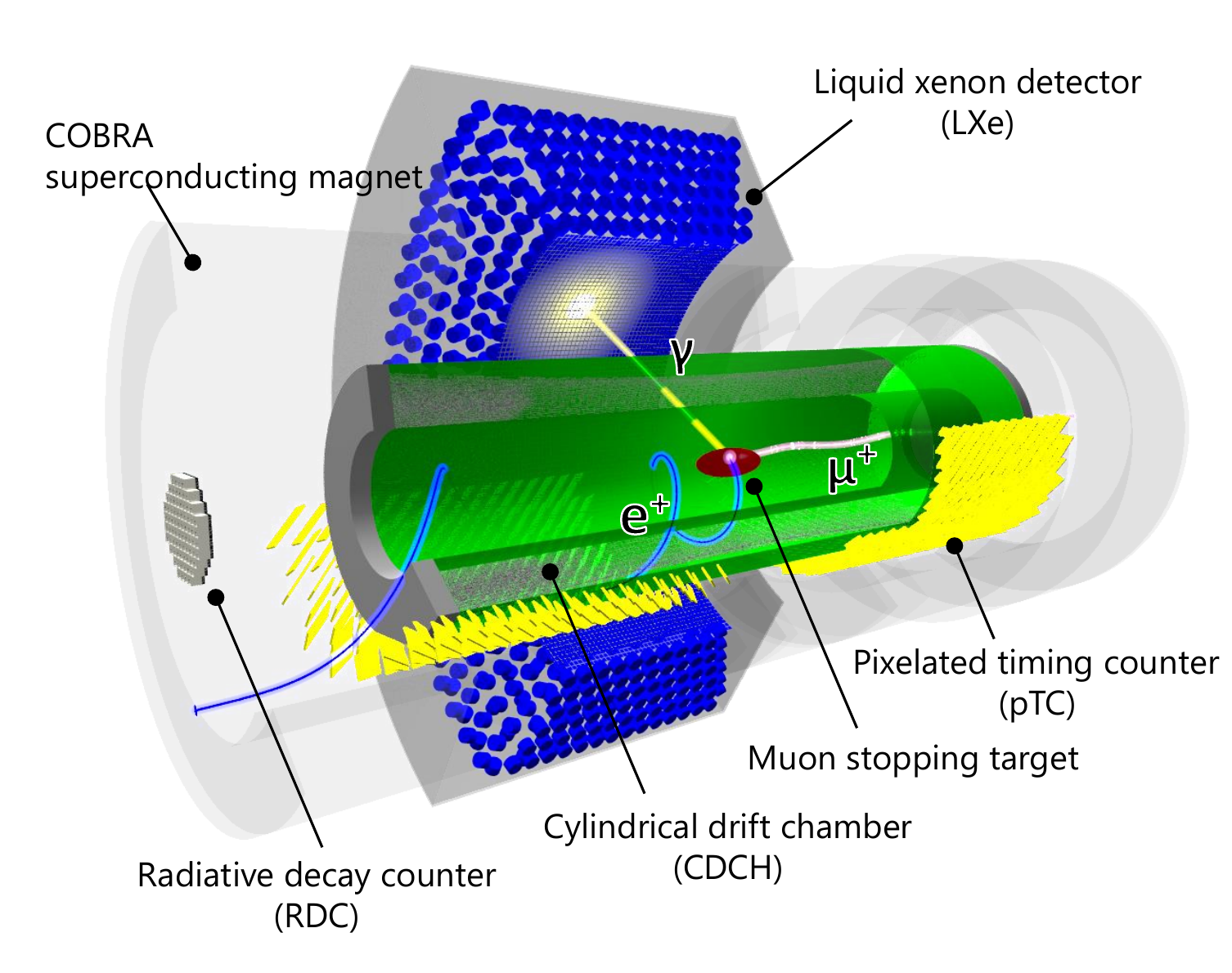}}
\end{minipage}
\hfill
\begin{minipage}{0.45\linewidth}
\centerline{\includegraphics[width=7cm]{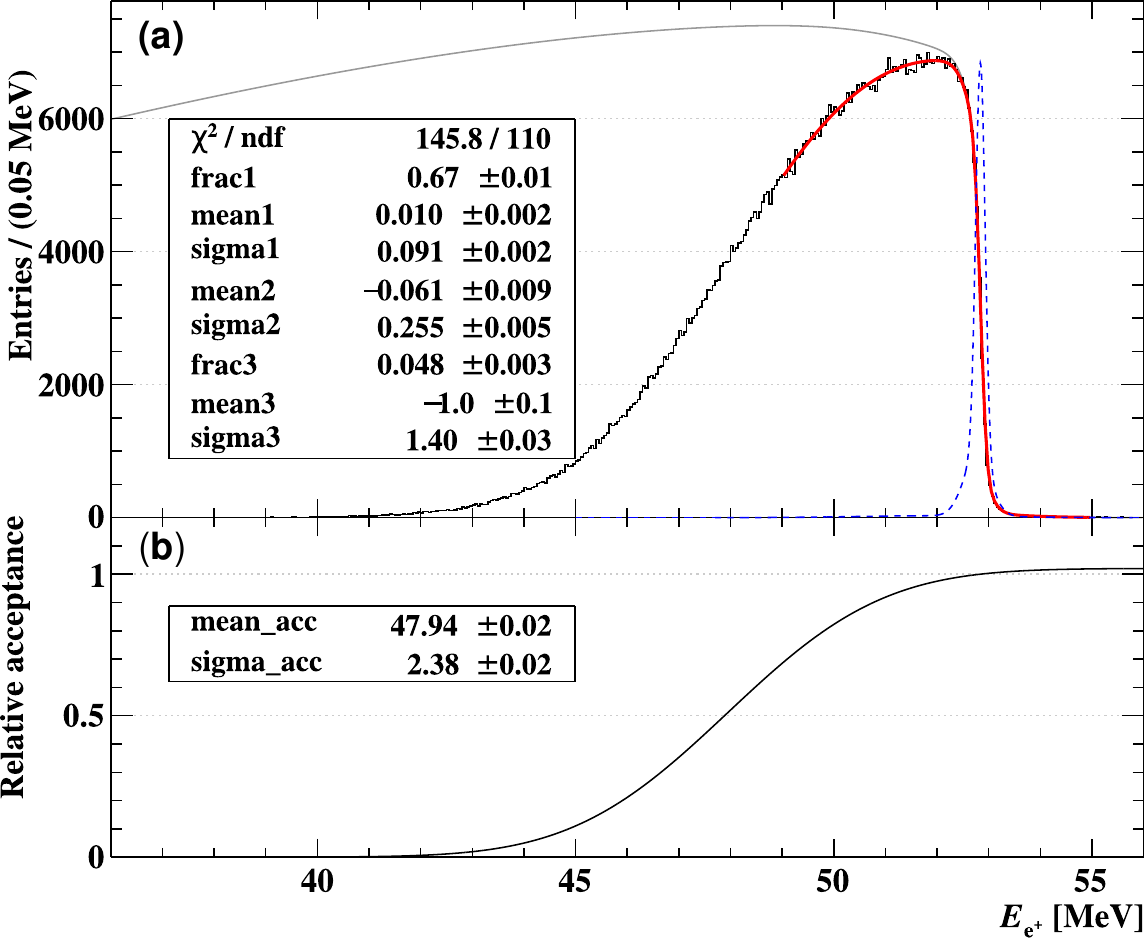}}
\end{minipage}
\caption[]{The MEG II detector apparatus\cite{MEGIIDesign} (left). (a) The fitted reconstructed positron momentum distribution on a linear scale, and (b) the fitted acceptance function. This yields the momentum scale, the momentum resolution and the expected number of background around the signal region\cite{MEGIIDetector} (right)}
\label{MEGIIAssm}
\end{figure}

The experiment starts with a nearly continuous $\mu^{+}$ beam with a rate of $3-5\cdot 10^{7} \mu/s$. The $\mu^{+}$ intersect a thin film target; about $\sim 85\% $ of the $\mu^{+}$ stop and decay in the target. This results in a flux of positrons, photons, and neutrinos. The detector apparatus is designed to detect the outgoing positron and photons; shown in Figure \ref{MEGIIAssm} (left). The basic principle is to detect the positrons in a magnetic spectrometer; the positron time is then measured in a set of pixelated timing counters. The photon is reconstructed in a liquid Xenon calorimeter. Offline, the positron is then propagated back to the stopping target using a Kalman filter; the photon is then projected back to the positron vertex at the target. The physics analysis then compares the relative time of the two particles ($t_{e\gamma}$), the relative angles of the particles ($\phi_{e\gamma}, \theta_{e\gamma}$), the positron momentum ($p_{e}$), and the photon energy ($E_{\gamma}$) to distinguish between the signal signature and the background. As mentioned, the dominant background is an accidental time coincidence; the number of accidental time coincidences ($N_{ACC}$) will scale like $N_{ACC} \propto R_{\mu}^{2} \cdot T \cdot \sigma_{ E_{\gamma}}^{2} \cdot \sigma_{p_{e}} \cdot \sigma_{t_{e\gamma}} \cdot \sigma_{\phi_{e\gamma}} \cdot \sigma_{\theta_{e\gamma}}$.\cite{MEGIIDesign} Therefore, it is critical to achieve the best resolutions to maximize the background rejection and thus achieve the best sensitivity.

The MEG II COBRA (constant bending radius magnet) magnet has a maximum field of 1.3 T at the stopping target and drops in field strength along the beam axis. The field forces the signal-like $e^{+}$ (52.8 MeV) away from the target upstream or downstream along the beam axis in a helical trajectory. The cylindrical drift chamber (CDCH) is an ultra-light single-volume stereo wire chamber designed to reconstruct the positrons. The wire chamber consists of approximately $1000$ 5 mm x 5 mm drift cells made of a central sense wire surrounded by a series of cathode/guard wires. The stereo geometry allows for precise reconstruction of the positron's position along the wire axis. The open-cell single volume detector design is significantly lighter than that of the MEG experiment in order to suppress multiple scattering contributions. The positron time is measured in a pixelated timing counter array (pTC) with 512 pixelated timing counters. Each counter is made of scintillator with SiPM readout on the two ends. The key upgrade with respect to its MEG counterpart is a higher hit multiplicity, with a signal-like $e^{+}$ intersecting an average of 9 counters. Each counter has a typical resolution of $\sim 100$ ps therefore resulting in $\sigma_{t_{e}}\sim 35$ ps. The fitted track with information from the CDCH and SPX is propagated back to the stopping target for measurements of the positron's position, direction, momentum, and time at the target.

The photon kinematics (time, position, energy) are measured in a single-volume $\sim800$ L liquid Xenon calorimeter (LXe). The upgraded MEG II calorimeter replaces the photo-multiplier tubes (PMTs) on the MEG calorimeter's inner face with 4096 multi-pixel photon counters (MPPCs). The high granularity of the MPPCs allows for significantly improved position resolution and reduces the light yield that escaped from the inner face in-between PMTs in MEG. 

In addition, the experiment has added a pair of cameras to photographically monitor the position, rotation, and deformation of the stopping target to suppress biases in the measured positron track angle at the target \cite{MEGIIDetector}. The experiment has also added a radiative decay counter (RDC) to eliminate some of the accidental time coincidences \cite{MEGIIDetector}.

\vspace{-0.3cm}
\subsection{Physics Result}
The MEG II collaboration started collecting physics data in 2021. A large scale effort was taken to calibrate and align the detectors to achieve the optimal resolutions. The physics result also required measurements of the kinematic resolutions to build the probability density functions (PDFs) for the likelihood analysis. The alignment, calibrations, and resolutions have been presented\^{}  \cite{MEGIIDetector}$^{,}$  \cite{CDCH}. As an example, in Figure \ref{MEGIIAssm} (right), we show the reconstructed $p_{e^{+}}$ distribution. The histogram is fit to the Michel $e^{+}$ distribution smeared with a triple gaussian (the resolution function) and an acceptance function (shown in subplot (b)). Effectively, the steeper the edge, the better the resolution. This is one of the critical measurements used to derive data-driven measurements of the positron momentum resolution, the number of expected accidental time-coincidences in the signal region, and the acceptance of the positrons as a function of momentum. 

\begin{figure}
\begin{minipage}{0.45\linewidth}
\centerline{\includegraphics[width=6cm]{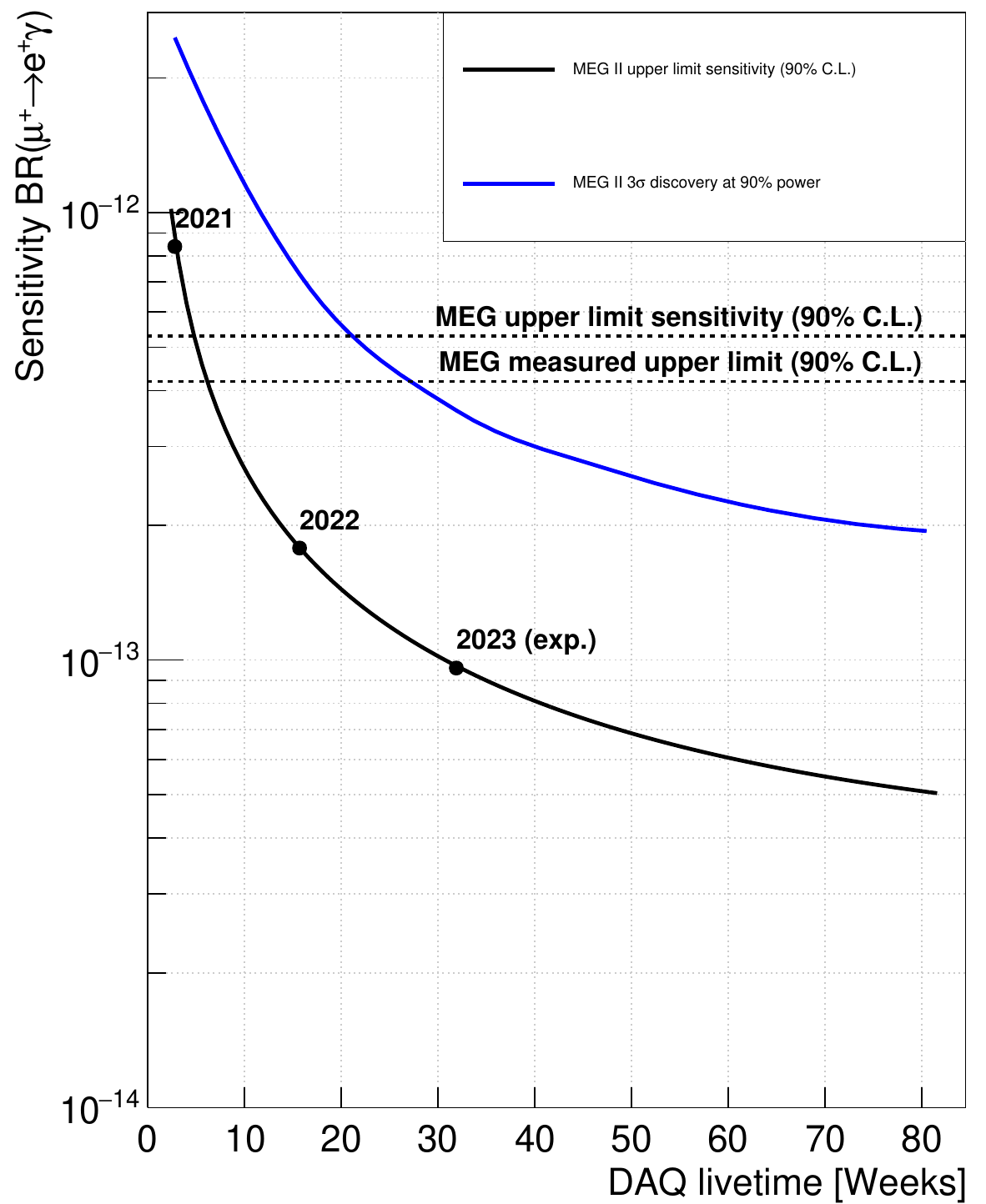}}
\end{minipage}
\hfill
\begin{minipage}{0.45\linewidth}
\centerline{\includegraphics[width=3.75cm]{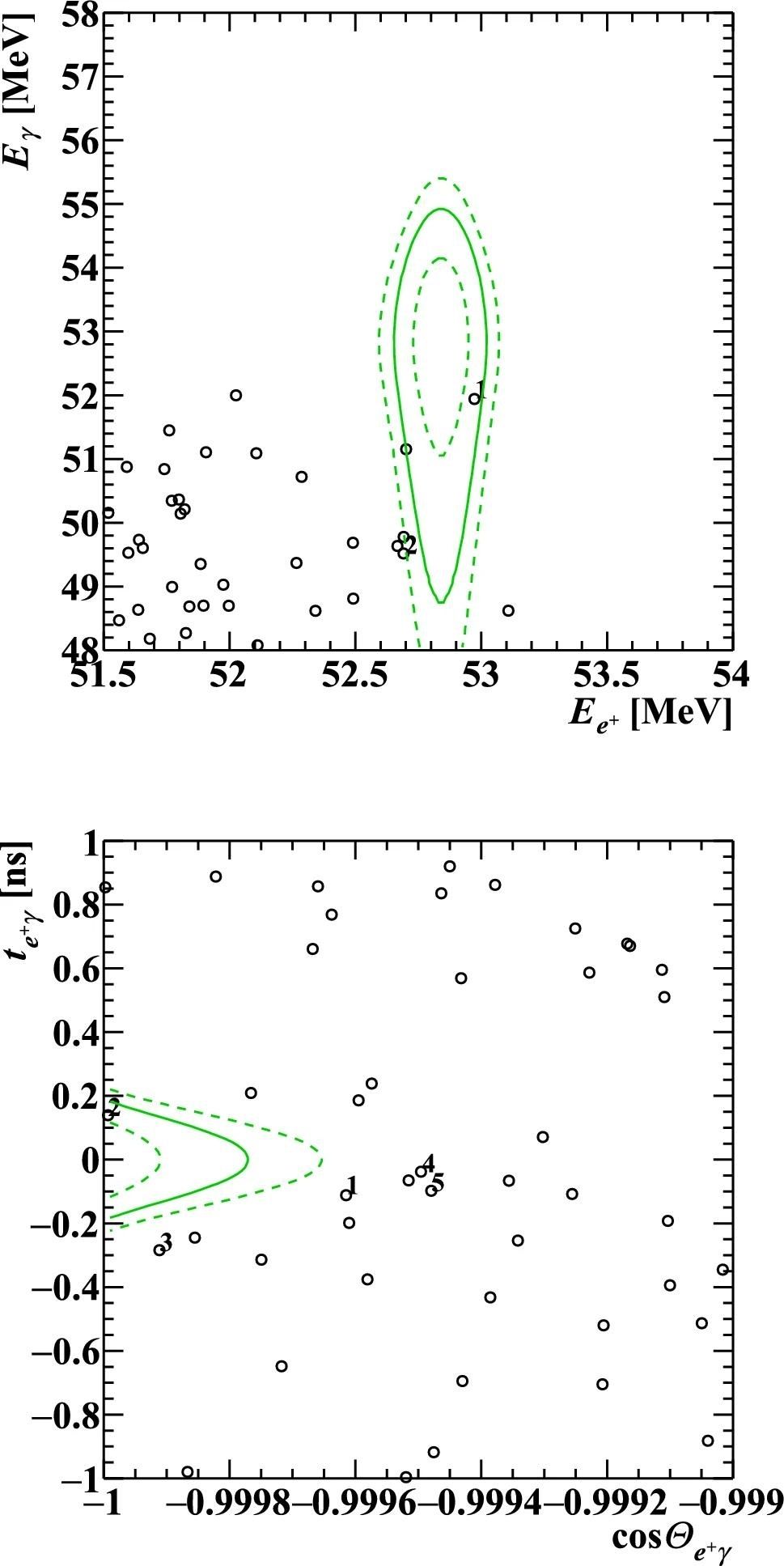}}
\end{minipage}
\caption[]{The expected sensitivity of the MEG II experiment over its full lifetime\cite{MEGIIDetector} (left). Example 2D slices of the kinematic phase space around the signal region in the 2021 data sample\cite{MEGIIPhysics}. (right)}
\label{fig:MEGSensitive}
\end{figure}

The sensitivity for the 2021 dataset is shown in Figure \ref{fig:MEGSensitive} (left); this sensitivity is the median limit from an ensemble of Monte Carlo experiments in the absence of signal using a blind maximum likelihood analysis. This Monte Carlo sensitivity measurement incorporates data-driven measurements of the kinematic resolutions of the signal and the expected number of background events from sideband studies such as the $p_{e}$ distribution described above. The sensitivity is then projected to the full MEG II dataset using the expected amount of data per year. The collaboration expects to be approaching its goal sensitivity with the data that has already been taken. This 2021 result shows that the MEG II detector apparatus is capable of handling the muon beam and the reconstruction is capable of achieving the resolutions required to reach its goal sensitivity of $6\cdot 10^{-14}$ at the 90\% confidence level. 

The MEG II collaboration applied a blind maximum likelihood analysis to the 2021 dataset. Unblinding the signal region did not result in any events near the signal signature and a limit of $7.5 \cdot 10^{-13}$ at the 90\% confidence level; this is consistent with the Monte Carlo sensitivity. Example kinematic phase space plots around the signal signature are shown in Figure \ref{fig:MEGSensitive} (right). 
 
\vspace{-0.3cm}
\subsection{Future Experimental Outlook}
A $\mu^{+} \rightarrow e^{+} \gamma$ experiment beyond MEG II will face many challenges. The designs will likely take advantage of beam rates of $>=10^{9} \mu/s$ available at several facilities. As the number of accidental time coincidences scales with the beam rate squared, a design similar to MEG II and a beam rate of $>= 10^{9} \mu/s$ would result in unacceptable background levels. An example design on the photon side is to instead convert the photons and then reconstruct the $e^{+}e^{-}$ pairs  \cite{Voena}$^{,}$\cite{Cavoto}. This comes at a significant loss in efficiency, but significantly improved photon energy resolution. Note, the number of accidentals scales with the photon energy resolution squared as the number of signal-energy photons drops off linearly with energy. In addition at $5\cdot 10^{7} \mu/s$, the CDCH already has a reduction of resolution from pileup ionization sites from out-of-time tracks intersecting the same cells; this is due to long drift times reaching $200-300$ ns and a high hit rate. This suggests an alternate approach to measuring the positrons is required. For example, a pixelated detector similar to that of Mu3e has been discussed \cite{Voena}. 
\vspace{-0.2cm}

\section{$\mu^{+} \rightarrow e^{+} e^{-} e^{+}$}
\vspace{-0.3cm}

\subsection{Experimental History and Timeline}
The current experimental limit on $\mu^{+} \rightarrow e^{+} e^{-} e^{+}$ is $1.0\cdot 10^{-12}$ at the 90\% confidence level and was set by SINDRUM at PSI in 1988 \cite{SINDRUM}. The upcoming Mu3e experiment at PSI is planning to surpass that, reaching $10^{-16}$ sensitivity \cite{Mu3e}. The experiment is actively preparing for commissioning with plans of an engineering run in 2025 and physics data-taking in 2025 or 2026. This section will focus on the phase I of the experiment, which aims for a $10^{-15}$ sensitivity result. 

\vspace{-0.3cm}
\subsection{Signal and Background}
The signal signature is a $e^{+} e^{-} e^{+}$ triplet consistent with a $\mu^{+}$ decay at rest in the stopping target. In the signal $e^{+} e^{-} e^{+}$ triplet, the three decay products are time coincident and position coincident at the stopping target, and conserve momentum. The backgrounds are analogous to the MEG II experiment. The first background is from $\mu^{+} \rightarrow e^{+} e^{-} e^{+} \nu_{e} \bar{\nu_{\mu}}$ decays (analogous to the RMD decay for MEG II). The second type of background is due to accidental time coincidences between two or more Standard Model processes.

\vspace{-0.3cm}
\subsection{Experimental Design}
\begin{figure}
\begin{minipage}{0.30\linewidth}
\centerline{\includegraphics[width=3.cm]{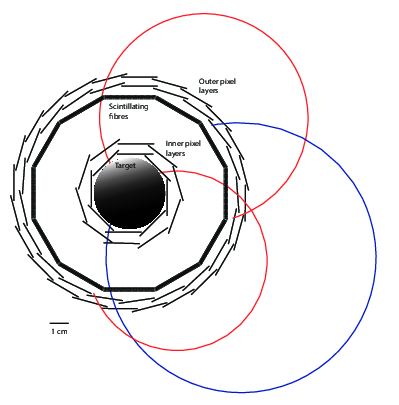}}
\end{minipage}
\hfill
\begin{minipage}{0.60\linewidth}
\centerline{\includegraphics[width=11cm]{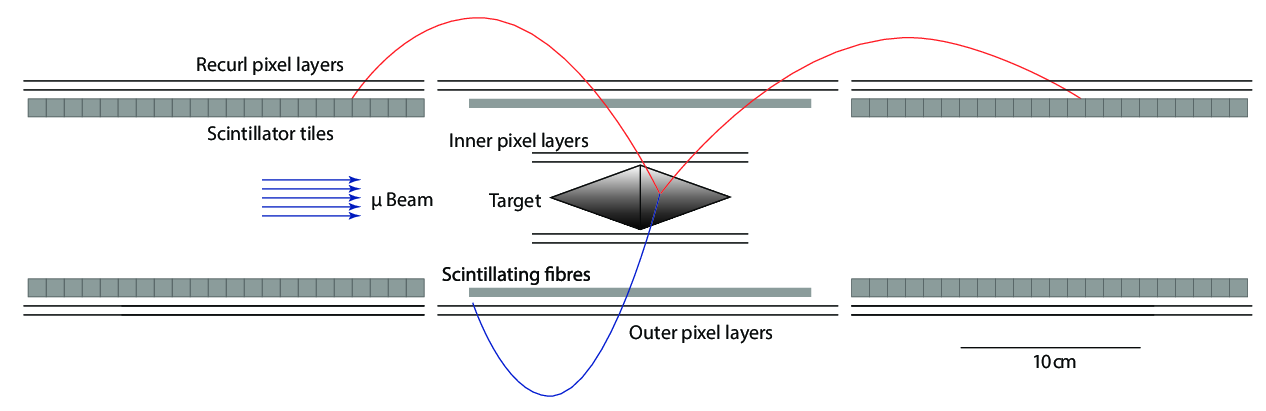}}
\end{minipage}
\caption[]{The Mu3e experimental design along the beam axis (left). The Mu3e experimental design perpendicular to the beam axis (right)}
\label{fig:mu3e}
\end{figure}

In phase I, Mu3e will use a $\mu^{+}$ rate of $10^{8} \mu/s$. The experiment starts with the $\mu^{+}$ stopping and decaying in a hollow double-cone target. The detector design is shown in Figure \ref{fig:mu3e}. Similar to MEG II, the experiment implements a magnetic field to reconstruct the $e^{+} e^{-}$ kinematics. The basic principle is to reconstruct $e^{+}$ and $e^{-}$ tracks to search for the signal signature $e^{+} e^{-} e^{+}$ triplets. The tracks are measured using a series of pixel sensors to reconstruct the track position and scintillating fibers and tiles to reconstruct the track time. These measurements can then be projected back to the target.  
At the core of the detector design is a set of 2844 Monolithic active pixel sensors (HV-MAPS)
specifically made for Mu3e (MUPIX) \cite{MuPIX}. The pixels has been made thin and small to yield precise position measurements along the $e^{+} e^{-}$ trajectories and to minimize scattering. The pixels consist of $50 \mu m$ of active material and $50 \mu m$ of support structure with a $80$x$80 \mu m^{2}$ size \cite{Mu3eBerger}. These pixel detectors are nearby precise timing devices (scintillating fibers and tiles). Combining the two measurements offers a precise position and timing measurement at multiple positions along the $e^{+} e^{-}$ trajectory. 

After leaving the target, the $e^{+} e^{-}$ immediately intersect "inner" pixel layers and then a set of scintillating fibers for an initial time estimate at the pixels. Either upstream or downstream of the target, the $e^{+} e^{-}$ "recurl" and intersect another pixel array and an array of scintillating tiles for the final precise timing measurement.

\vspace{-0.3cm}

\subsection{Future Experimental Outlook}
In the Mu3e phase II, the experiment plans to reach a sensitivity of $2 \cdot 10^{-16}$ at the 90\% confidence level by using a higher beam rate of $2\cdot 10^{9} \mu/s$ and adding additional pixel recurl layers and scintillator layers to aid in the combinatoric issues at even higher beam rates and to improve the resolutions (e.g. momentum). 
\vspace{-0.3cm}

\section{$\mu^{-} N \rightarrow e^{-} N$}
\vspace{-0.3cm}
\subsection{Experimental History and Timeline}
The current best limit on the $\mu^{-} N \rightarrow e^{-} N$ conversion is $7\cdot 10^{-13}$ at the 90\% confidence level in a gold target \cite{SINDRUMII}; this limit was set by SINDRUM II at PSI in 2006. There are two upcoming experiments searching for $\mu^{-} N \rightarrow e^{-} N$, both intending to reach $<10^{-16}$ sensitivity (Mu2e and COMET Phase II). 

The first is the Mu2e experiment \cite{Mu2e} at Fermilab, which is currently in the construction phase. The collaboration plans to bring all detectors and magnets into the Mu2e hall for cosmic data-taking starting in 2025. Mu2e will have the required beam for the full 2027 calendar year for commissioning and physics data-taking. 

The second $\mu^{-} N \rightarrow e^{-} N$ experiment is the COMET experiment at J-PARC \cite{COMET}; the experiment is constructing phase I, which will plan to improve upon the limit set by SINDRUM II by two orders of magnitude. COMET phase I is planning on being physics ready in 2026. 

This section will focus on the Mu2e experiment. However, considerations for both the COMET and Mu2e experiments are very similar.

\begin{figure}[t]
\begin{minipage}{0.25\linewidth}
\centerline{\includegraphics[width=6cm]{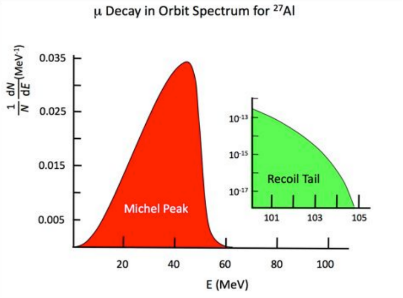}}
\end{minipage}
\hfill
\begin{minipage}{0.65\linewidth}
\centerline{\includegraphics[width=11cm]{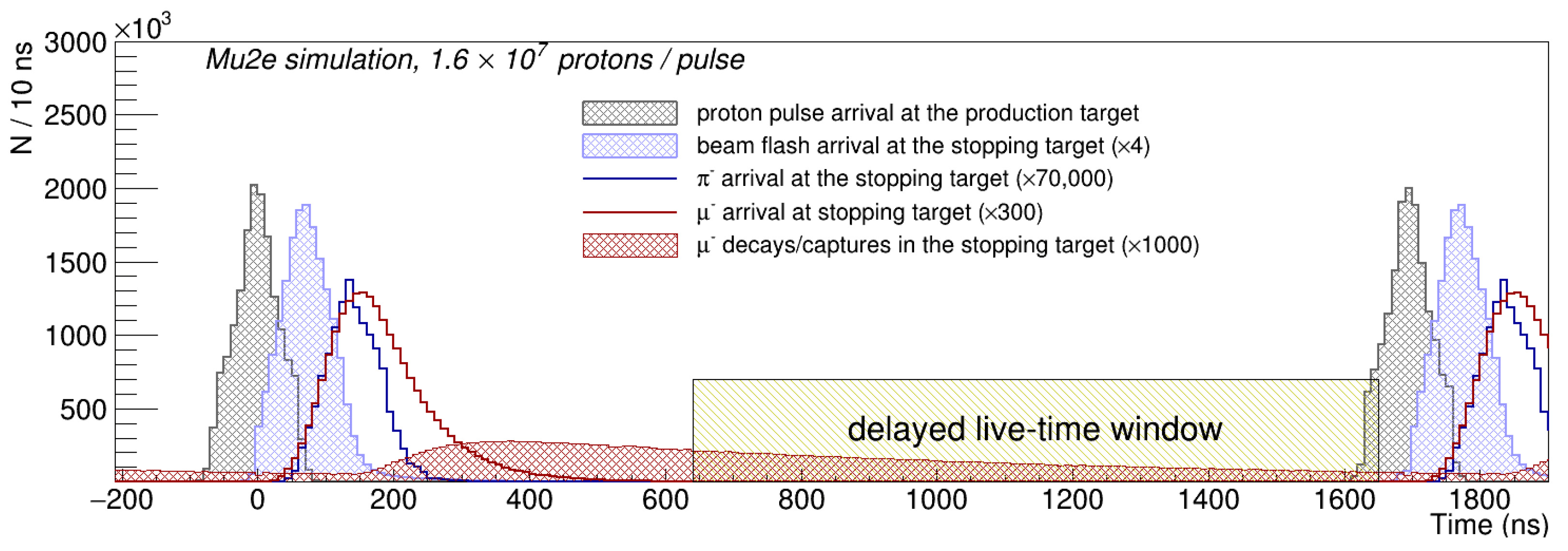}}
\end{minipage}
\caption[]{The $\mu^{-}$ DIO spectrum and the relevant recoil tail\cite{Mu2e} (left). The pulsed Mu2e beam structure\cite{Mu2eUni} (right)}
\label{fig:Mu2eDIO}
\end{figure}

\vspace{-0.3cm}
\subsection{Signal and Background}
The signal signature is the conversion of a muon in the field of the target nucleus. There is a coherent recoil of the muon off the target nucleus; this results in an electron with a monochromatic energy. This signal energy is nearly the muon rest mass with small corrections for binding energy and the recoil energy of the nucleus, both of which are dependent on the nucleus: $E_{e} =  \ m_{\mu}c^{2} - B_{\mu}(Z) - C(A)= 104.97$ MeV in Aluminum, the target nucleus for Mu2e. 

To achieve the $10^{-16}$ sensitivity, the experiment must have extreme background suppression. Over the experiment's full lifespan, the Mu2e experiment expects $\sim 10^{18}$ stopped muons and expects to reduce the background to a combined expected $0.41 \pm 0.03$ events over the full dataset \cite{Mu2eBern}.  Below, we detail the major experimental backgrounds.

Similar to both the MEG II and Mu3e experiments, there is background due to a Standard Model process that results in an electron near the signal energy and two low momentum neutrinos. Here, a muon can decay in orbit (DIO) of the nucleus, which distorts the "clean" $p_{e^{+}}$ distribution shown in Figure \ref{MEGIIAssm} (right) as the $e^{-}$ recoils off the nucleus. DIO spectrum is shown in Figure \ref{fig:Mu2eDIO} (left). Near the DIO high momentum limit, this is only distinguishable from the signal via a precise momentum measurement, which largely motivates the tracker design. 

The second background comes from $\pi^{-}$ used to produce the $\mu^{-}$ (the beamline will be discussed in the next subsection). If the $\pi^{-}$ intersects the $\mu$ target it can produce a high energy photon ($\pi^{-} N \rightarrow \gamma N_{*}$) that can then produce an electron at the signal energy. This is known as radiative pion capture (RPC). This background motivates a pulsed beam structure, which will be discussed below in detail.  

The other major background is from cosmic rays. Any cosmic ray can intersect some material, which can break free an electron at the signal energy. This is indistinguishable from the signal (particularly if the cosmic intersects the target). Simulations suggests cosmic rays in the detector area will produce roughly one signal-like electron per-day. To eliminate this background, the entire detector region will be surrounded by a cosmic ray veto system (CRV). 

There are other expected backgrounds from antiprotons, muon decay in flight, etc., but these are not the dominant background and will not be discussed here. More information is shown here\cite{Mu2e}$^{,}$ \cite{Mu2eBern}$^{,}$ \cite{Mu2eUni}.

In summary, the Mu2e experiment requires precise momentum resolution to eliminate the DIO tail, requires a pulse beam to eliminate background caused from RPC pions that intersect the muon target, and requires a cosmic ray veto system surrounding the detector area to eliminate cosmic ray produced signal-like electrons. 
\vspace{-0.3cm}

\subsection{Experimental Design}

\begin{figure}[t]
\begin{minipage}{0.35\linewidth}
\centerline{\includegraphics[width=7.5cm]{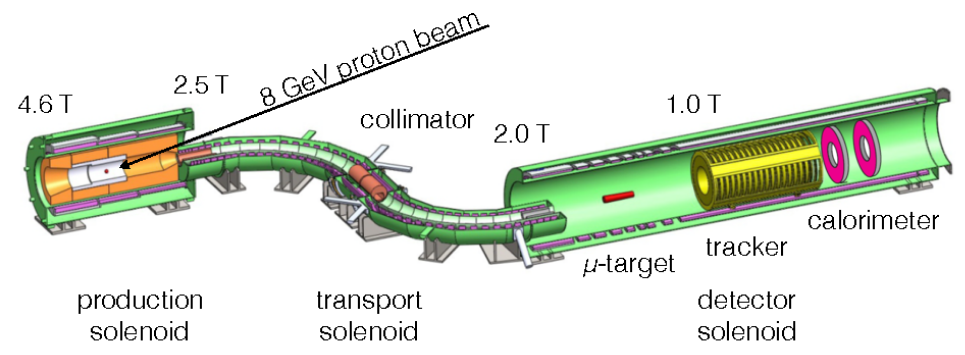}}
\end{minipage}
\hfill
\begin{minipage}{0.35\linewidth}
\centerline{\includegraphics[width=4.5cm]{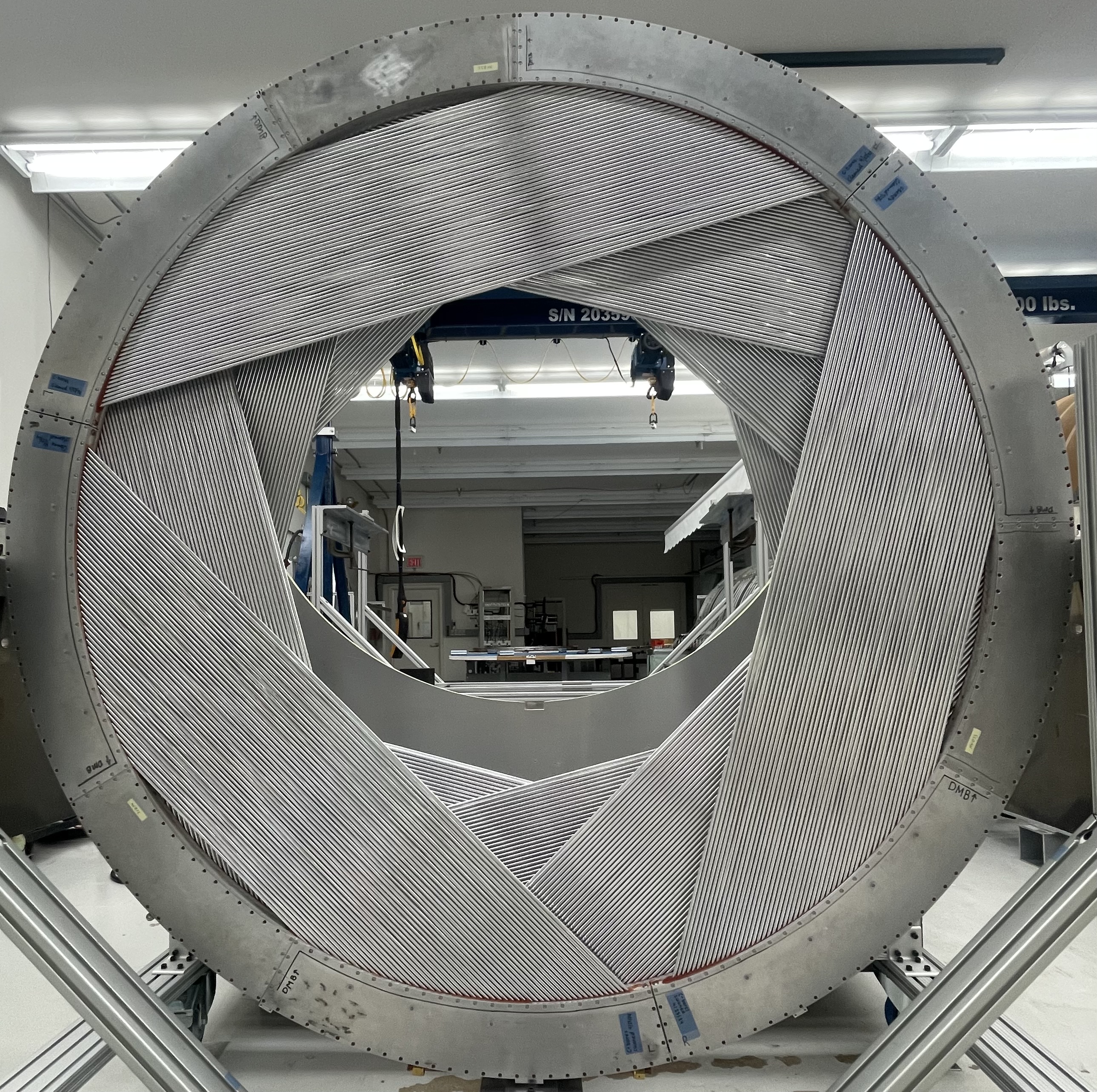}}
\end{minipage}
\hfill
\begin{minipage}{0.25\linewidth}
\centerline{\includegraphics[width=5.5cm]{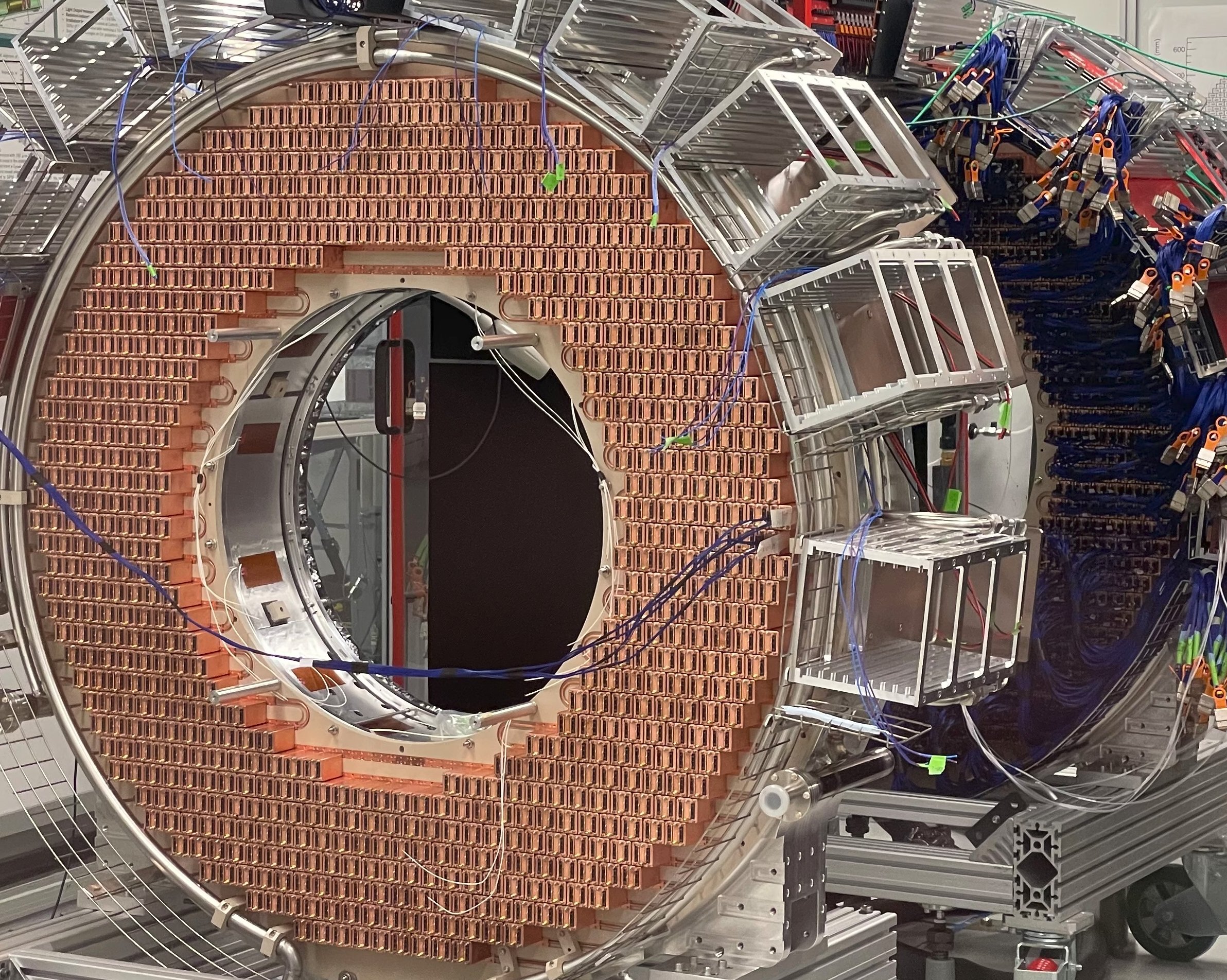}}
\end{minipage}
\caption[]{The Mu2e experimental design (left). A single Mu2e tracker station (1/18 of the tracker) (middle). The Mu2e calorimeter (right).}
\label{fig:Mu2eDetect}
\end{figure}
The Mu2e experimental design is shown in Figure \ref{fig:Mu2eDetect} (left). The experiment starts with the Fermilab proton beam intersecting a tungsten target inside the production solenoid. The production solenoid is a strong (4.6 T) magnetic field designed to remove a majority of the unwanted particles resulting from the proton-target interaction. The experiment collects the backward moving $\pi^{-}$ (backward with respect to the proton beam direction). This results in a flux of $\pi^{-}$ that decay to $\mu^{-}$ in flight. The transport solenoid "S" shape selects negative charge particles and the required low momentum muons. Finally, the $\mu^{-}$ arrive at the detector solenoid and stop in the Aluminum muon target. The detector solenoid forces the electrons on a helical trajectory and draws them towards the detector apparatus. 

Note, that unlike Mu3e or MEG II, this is a "pulsed" proton beam (1700 ns). The pulsed proton beam results in a pulse of pions; these pions have a significantly shorter lifetime than the muon lifetime in the Aluminum target($\tau_{\mu_{Al}}=864$ ns). To avoid the pion RPC background, the experiment waits until nearly all the pions have decayed to start collecting data, therefore eliminating a vast majority of the RPC background. The pulse timing is detailed in Figure \ref{fig:Mu2eDIO} (right). Note, this requires monitoring the "extinction" of the proton beam in between pulses. To achieve the desired background suppression, a proton beam extinction of $10^{-10}$ is required. This is measured by an independent extinction monitor.

The core Mu2e detector is a straw tube tracker designed for reconstructing electrons at the signal energy. Here, the geometry contains an inner hole to allow the typical approximately $50$ MeV electron (see Figure \ref{fig:Mu2eDIO} (left)) to escape without intersecting the detector volume. The detector contains a total of approximately $20$k 5 mm diameter straws that are positioned in $120^{\circ}$ "panels" each with 96 straws. These panels are oriented in circles (3x$120^{\circ}$) and along the beam axis. Figure \ref{fig:Mu2eDetect} (middle) shows one station made of 12 panels; the overlap of straws at varying angles allows for stereo reconstruction. The detector will consist of 18 stations oriented along the beam axis. The detector reconstructs the electron track momentum with an expected resolution of $\sim150$ keV; as a reminder, this measurement is used to distinguish between the signal and the DIO tail. 

After intersecting the tracker, the electrons intersect a calorimeter; it is shown in Figure \ref{fig:Mu2eDetect} (right). The calorimeter consists of two circular disks with a total of 674 undoped CsI crystals. The calorimeter is used for particle identification ($\mu$ vs. $e$), track time $T_{0}$ measurements, and an independent trigger. The detector region is additionally surrounded by a cosmic ray veto system to identify (and thus nearly eliminate) the cosmic ray induced background. The CRV consists of 4 layers of scintillation counters (5504 in total) with embedded fibers readout by SiPMs. 

\vspace{-0.3cm}

\subsection{Future Experimental Outlook}
At both Fermilab and J-PARC, experimental designs beyond Mu2e and COMET Phase II are already being investigated. Advanced searches for $\mu^{-} N \rightarrow e^{-} N$ can more easily take advantage of higher muon beam rates as the search does not have the time coincidence background like $\mu^{+} \rightarrow e^{+} \gamma$ and $\mu^{+} \rightarrow e^{+} e^{-} e^{+}$. An improved search likely requires a modified beam structure such as a muon storage ring to further reduce the background from the beam \cite{Kuno}$^{,}$ \cite{CLFVFermilab}. This could also result in a lower momentum muon beam therefore requiring less target material and therefore a more precise measurement of the electron momentum. 
\vspace{-0.3cm}

\section{Conclusion}
The proceedings reports on the status of CLFV experiments with muons. The MEG II experiment, searching for $\mu^{+} \rightarrow e^{+} \gamma$, published a result indicating it is on track to reach its goal of a 10x improvement in sensitivity beyond the MEG experiment. The Mu3e experiment searching for $\mu^{+} \rightarrow e^{+} e^{-} e^{+}$ plans to improve upon the current sensitivity limit by $10^{4}$ with data-taking planned to start in 1-2 years. The Mu2e and COMET experiments, both searching for $\mu^{-} N \rightarrow e^{-} N$, plan to improve upon the current sensitivity limit by $10^{4}$  with data-taking starting in 2027 and 2026 respectively.

\vspace{-0.3cm}

\section*{Acknowledgments}
\vspace{-0.2cm}

We are grateful for the vital contributions of the Fermilab staff and the technical staff of the participating institutions. This work was supported by the US Department of Energy; the Istituto Nazionale di Fisica Nucleare, Italy; the Science and Technology Facilities Council, UK; the Ministry of Education and Science, Russian Federation; the National Science Foundation, USA; the National Science Foundation, China; the Helmholtz Association, Germany; and the EU Horizon 2020 Research and Innovation Program under the Marie Sklodowska-Curie Grant Agreement Nos.  734303, 822185, 858199, 101003460, and 101006726. This document was prepared by members of the Mu2e Collaboration using the resources of the Fermi National Accelerator Laboratory (Fermilab), a U.S. Department of Energy, Office of Science, HEP User Facility. Fermilab is managed by Fermi Research Alliance, LLC (FRA), acting under Contract No. DE-AC02-07CH11359.

\end{document}